\documentclass[%
 reprint,
 amsmath,amssymb,
 aps,prb, twocolumn, showpacs
]{revtex4-1}

\usepackage{graphicx}
\usepackage{dcolumn}
\usepackage{bm}
\usepackage{float}
\usepackage{caption}
\usepackage{subfigure}
\usepackage{amsmath}
\usepackage{mathrsfs}
\usepackage{amssymb}
\usepackage{hyperref}
\usepackage{xcolor}

\begin{document}

\preprint{}

\title{Excitations of Atomic Vibrations in Amorphous Solids}

\author{Li Wan}
\email{lwan@wzu.edu.cn}
\affiliation{Department of Physics, Wenzhou University, Wenzhou 325035, P. R. China}
\date{\today}

\begin{abstract}
We study excitations of atomic vibrations in the reciprocal space for amorphous solids. There are two kinds of excitations we obtained, collective excitation and local excitation. The collective excitation is the collective vibration of atoms in the amorphous solids while the local excitation is stimulated locally by a single atom vibrating in the solids. We introduce a continuous wave vector for the study and transform the equations of atomic vibrations from the real space to the reciprocal space. We take the amorphous silicon as an example and calculate the structures of the excitations in the reciprocal space. Results show that an excitation is a wave packet composed of a collection of plane waves. We also find a periodical structure in the reciprocal space for the collective excitation with longitudinal vibrations, which is originated from the local order of the structure in the real space of the amorphous solid.
\begin{description}
\item[Keywords]
{atomic vibrations; amorphous solid; excitation; wave packet; local order }
\end{description}
\end{abstract}

\maketitle

\section{Introduction}
Atomic vibrations have been well understood in crystalline solids both theoretically and experimentally~\cite{born,ziman, dove}. In the crystalline solids, atoms vibrate around their equilibrium positions and excitations of the lattice vibrations are known as phonons. Each phonon has a well defined crystal wave vector and a vibration frequency. To get phonons, the equation of lattice dynamics need to be transformed from the real space to the reciprocal space. In the transformation, lattice plane waves are introduced, and play their roles as lattice Fourier transformations. By using the lattice periodicity in the crystalline solids, the lattice Fourier transformations diagonalize the dynamics matrix to get the phonons. One lattice plane wave in the transformation gets one kind of phonons. However, in the amorphous solids, the periodicity of the lattices is broken, which makes phonons have no definition in the solids~\cite{henry1, henry, henry2,phononfail1,phononfail2}. From the quantum mechanical view of point, the lack of the periodical lattices in the amorphous solids makes the crystal wave vector not a good quantum number any longer and the lattice Fourier transformations based on the crystal wave vector fail to diagonalize the Hamiltonian of the system. Such invalidity of the Fourier transformations breaks the bridge between the real space and the reciprocal space for the amorphous solids. Therefore, how to get excitations in the reciprocal space for the amorphous solids still remains as a problem.\\

To study atomic vibrations of the amorphous solids, various techniques have been applied. In the techniques, molecular dynamics (MD) simulation as a powerful tool has been widely used~\cite{MDFrenkel}. The MD simulation is to solve a set of equations of motion by the Newton's law for the positions and velocities of atoms in the amorphous solids. After a long time evolving, the atoms reach their equilibrium positions. Besides generating the amorphous solids, the MD technique has been widely used to study the thermal transport in the solids~\cite{MD1, MD2, MD3}. The calculations of the thermal conductivity by MD are based on the Green-Kubo formula. The informations of lattice dynamics obtained from MD simulations are in the real space and can not provide the structures of the excitations in the reciprocal space for the amorphous solids.\\

Besides the MD technique, lattice dynamics(LD) method is a different way to get the informations of the atomic vibrations~\cite{dove}. LD method has been generalized to the amorphous solids and is used to diagonalize the dynamics matrix to get frequency spectrum. The diagonalizing is conducted in the real space for the solids and the density of state then can be obtained. To make the results close to the real amorphous solids, the LD method normally is applied on a supercell. Philip Allen, Joseph Feldman and others conducted the LD calculations on a supercell of amorphous silicon~\cite{LD1,LD2,LD3}. Basing on the calculations, they classified the vibrational states into three categories: propagons, diffusons and locons. Propagons occupy the bottom of the frequency spectrum and are considered to be delocalized. Propagons can propagate through out the whole disorder system and behave like phonons in crystalline solids. Locons are high frequency modes and considered to be spatially localized. The modes termed with diffusons have the frequencies in the range between the propagons and the locons. Diffusons are not spatially localized and they contribute to the thermal transport through diffusive processes rather than the propagation of the propagons. Such taxonomy introduced by Allen, Feldman and their colleagues opened up a new perspective on vibrational modes in amorphous solids and provides insights into the thermal transport in the solids~\cite{LD4}.\\

Similar to the MD simulation, LD method identifies the vibration modes not in the reciprocal space. The wave vector is not involved in the LD method, which still can not provide the excitations in the reciprocal space. The propagons, diffusons and locons are not the elementary excitations in the amorphous solids. The classifying of the propagons, diffusons and locons with clear range boundaries is still under debates. To figure out the informations of atomic vibrations from the reciprocal space, Moon et.al defined a dynamic structure factor to study the crossover frequency from the propagating excitations to diffusive vibrations~\cite{moon1, moon2}. In their work, a supercell is considered to be periodical and vibration frequencies used for the dynamic structure factor are all at $\Gamma$. The relation between the wave vector and the frequency discovered by Moon et.al is not the dispersion relation of elementary excitations. Similar work is performed by Seyf and Henry by defining various versions of the structure factor~\cite{LD4}.\\

In this study, we transform the equations of the atomic vibrations from the real space to the reciprocal space and provide the excitations in the reciprocal space for amorphous solids. The transformations are rigorous. The wave vector involved in the transformations can cover the whole reciprocal space. And the dispersion relation of the excitations can be clearly shown in the reciprocal space, which paves a way for the study of thermal transport in amorphous solids.\\

\section{theory}
\label{theory}
We consider an amorphous solid of three dimensional structure. The total number of atoms in the solid is $N$. Each atom in the solid is vibrating around its own equilibrium position. For the $l$-th atom with the mass of $M_l$, the equilibrium position is denoted by $\vec{R}_l$ and the displacement away from $\vec{R}_l$ for the vibration is by $\vec{r}_l$. We fix a rectangle coordinate system in the solid with the three axises denoted by $x$, $y$ and $z$ respectively. Due to the random arrangement of atoms in the amorphous solid, it is meaningless to specify concrete directions for the axises like what can be done in crystalline solids. We use the coordinate system only to denote the components of vectors conveniently. The component of $\vec{R}_l$ along a coordinate axis, say $x$ axis, is denoted by $R_{l,x}$. Similar notation is applied on $\vec{r}$.\\

In the vibration, each atom follows the dynamics equation of 
\begin{equation}
\label{dynequ}
M_l\ddot{r}_{l,\alpha}=-\sum_{p,\beta}\frac{\partial^2 \Psi}{\partial R_{l,\alpha}\partial R_{p,\beta}} r_{p,\beta}.
\end{equation} 
In the above equation, the double dots on the top of $r$ means the second order derivative of $r$ with respect to time $t$. The total potential energy of the amorphous solid is denoted by $\Psi$, which is functional of equilibrium positions $\vec{R}$ of atoms. In the subscripts, $l$ and $p$ are the indexes to label atoms. $\alpha$ and $\beta$ both are the coordinates $x$, $y$ or $z$. The second order partial derivative of $\Psi$ with respect to $R_{l,\alpha}$ and $R_{p,\beta}$ actually is the force constant between the $l$-th atom and the $p$-th atom when the $l$-th atom moves along $\alpha$ direction and the $p$-th atom along $\beta$ direction. For convenience, we introduce  $\Phi_{l,p}^{\alpha, \beta}=(\frac{\partial^2 \Psi}{\partial R_{l,\alpha}\partial R_{p,\beta}})(\frac{1}{\sqrt{M_lM_p}})$ for the force constant. We also introduce a quantity $\vec{u}_l=\sqrt{M_l}\vec{r}_l$ to simplify the dynamics equation (\ref{dynequ}) as
\begin{equation}
\label{dynequreduced}
\ddot{u}_{l,\alpha}=-\sum_{p,\beta}\Phi_{l,p}^{\alpha, \beta} u_{p,\beta},
\end{equation}
which is general for solids consisting of multi-species with various masses.\\

Our goal is to transform Eq.(\ref{dynequreduced}) from the real space to the reciprocal space to get excitations. It is known that the lattice Fourier transformation that has been widely used in the crystalline solids to get phonons now is invalid in the amorphous solid due to the lack of the lattice periodicity. Thus, we need a transformation different to the lattice Fourier transformation for our goal. Before we introduce the transformation to achieve our goal, we note that there exist two kinds of excitations in the amorphous solid, collective excitation and local excitation. The collective excitation is for the collective vibrations of the atoms, while the local excitation is stimulated by a single atom. The local excitation is spatially localized around the simulating atom and decays its intensity in propagating away from the atom. The decaying of the local excitation is due to the random scattering by the other atoms around the stimulating atom. We will introduce two transformations on Eq.(\ref{dynequreduced}) for the collective excitation and the local excitation respectively in the following.\\

\subsection{Collective Excitation} 
We start from Eq.(\ref{dynequreduced}), and introduce a wave vector $\vec{\kappa}$ for a transformation. The wave vector $\vec{\kappa}$ is continuous and is different from the crystal wave vector that is discrete for the crystalline solids. We denote the imaginary unit by $i$ and define a transformation for an normal coordinate $Q_{\vec{\kappa}}^{\alpha}=(1/\sqrt{N})\sum_l e^{-i\vec{\kappa}\cdot \vec{R}_l}u_{l,\alpha}$ in the spirit of Fourier transformation. The wave vector $\vec{\kappa}$ is continuous because of the absence of the lattice periodicity in the amorphous solid, and ranges from $-\infty$ to $+\infty$ along any direction. The collective excitation(CE) is the collective vibration of all the atoms and propagate without decaying its intensity. Thus, the wave vector $\vec{\kappa}$ introduced for the CE must be real. Or, the imaginary part of the wave vector will decay the intensity of the CE. The complex wave vector will be applied for the local excitation, but not for the CE.\\

Physically, the normal coordinate $Q_{\vec{\kappa}}$ defines a plane wave with the wave vector of $\vec{\kappa}$. The term of $u_{l,\alpha}e^{-i\vec{\kappa}\cdot \vec{R}_l}$ in $Q_{\vec{\kappa}}$ shows the magnitude and phase of the plane wave at the equilibrium position of the $l$-th atom. In the crystalline solids, such plane wave defines a phonon. The frequency of the phonon corresponds to the wave vector through the dispersion relation. All possible solutions to the dispersion relation are curves for the crystalline solids. However, it is not the case in the amorphous solid, where the curves in the dispersion relation are reorganized, broadening and even dispersive. For a given wave vector $\vec{\kappa}$, we multiple the both sides of Eq.(\ref{dynequreduced}) by $(1/\sqrt{N})e^{-i\vec{\kappa}\cdot \vec{R}_l}$ and sum the both sides over the total atoms. Then we get a new equation, reading
\begin{align}
\label{CEequ}
\ddot{Q}_{\vec{\kappa}}^{\alpha}=-\sum_{\beta}\frac{1}{V_{\vec{\kappa}'}}\int F_{\vec{\kappa},\vec{\kappa}'}^{\alpha, \beta}Q_{\vec{\kappa}'}^{\beta}d\vec{\kappa}'
\end{align}
with $F_{\vec{\kappa},\vec{\kappa}'}^{\alpha, \beta}=\sum_{l,p}e^{-i\vec{\kappa}\cdot \vec{R}_l}\Phi_{l,p}^{\alpha, \beta}e^{i\vec{\kappa}'\cdot \vec{R}_p}$. On the left hand side of Eq.(\ref{CEequ}), we have used the definition of the normal coordinate $Q$ and kept the second order derivative with respect to time. On the right hand side of Eq.(\ref{CEequ}), $V_{\vec{\kappa}'}$ is the volume for the integration $\int d\vec{\kappa}'$ in the reciprocal space. The details for the derivation of Eq.(\ref{CEequ}) has been shown in Appendix A. Now we have transformed the dynamics equation from the real space to the reciprocal space by Eq.(\ref{CEequ}).\\

To go further, we express Eq.(\ref{CEequ}) in matrix form. We arrange $Q_{\vec{\kappa}}^{\alpha}$ in one column as a vector $Q$. Each entry in $Q$ is indexed by both of $\vec{\kappa}$ and $\alpha$. We arrange the force constant $\Phi_{l,p}^{\alpha, \beta}$ in the dynamics matrix $\Phi$. The row of $\Phi$ is indexed by both of $l$ and $\alpha$ while the column of $\Phi$ is by $p$ and $\beta$. Since $\alpha$ or $\beta$ represents three perpendicular directions ($x$, $y$ and $z$), $\Phi_{l,p}^{\alpha, \beta}$ is a $3\times 3$ sub-matrix by varying  $\alpha$ and $\beta$ for a given pair of $l$ and $p$. Then, we define a matrix $\xi$ for $e^{-i \vec{\kappa}\cdot \vec{R}_l}$. The row of $\xi$ is indexed by both of $\vec{\kappa}$ and $\beta$ while the column is indexed by both of $l$ and $\alpha$, even though the indexes $\alpha$ and $\beta$ do not appear in the element $e^{-i \vec{\kappa}\cdot \vec{R}_l}$ of $\xi$. By varying the indexes of $\alpha$ and $\beta$ for each given pair of row index $\vec{\kappa}$ and column index $l$ in $\xi$, we have a $3\times 3$ sub-matrix. The sub-matrix in $\xi$ is a $3\times 3$ identity matrix times $e^{-i \vec{\kappa}\cdot \vec{R}_l}$. In this way, the sizes of the matrices are consistent for the matrix product and Eq.(\ref{CEequ}) is still hold in the matrix form. According to Eq.(\ref{CEequ}), $\xi^{\dagger}$ is in between $\Phi$ and $Q$. Then, we discrete the integral of Eq.(\ref{CEequ}). We set the infinitesimal volume $\Delta \vec{\kappa}'$ for the reciprocal space and divide the total volume $V_{\vec{\kappa}'}$ by $\Delta \vec{\kappa}'$ to get the total discrete number $\mathcal{N}$ for the wave vector. Finally, we introduce a matrix $F=\frac{1}{\mathcal{N}}\xi\cdot \Phi \cdot \xi^{\dagger}$ for the component $(1/V_{\vec{\kappa}'})F_{\vec{\kappa},\vec{\kappa}'}^{\alpha,\beta}$. Eq.(\ref{CEequ}) then is expressed in the matrix form, reading
\begin{align}
\label{CEequQ}
\ddot{Q}=-\frac{1}{\mathcal{N}}\xi\cdot \Phi \cdot \xi ^{\dagger}\cdot Q=-F \cdot Q,
\end{align}
which is equivalent to Eq.(\ref{CEequ}) in the limit of $\Delta \vec{\kappa}' $ approaching zero. The dot between two matrices represents the matrix product.\\

It could be found that matrix $F$ is not diagonal, meaning that the plane waves $u_{l,\alpha}e^{-i\vec{\kappa}\cdot \vec{R}_l}$ with different wave vectors $\vec{\kappa}$ interact with each other. It is the nature of the disorder system, comparing to the case of crystalline solids in which $F$ is diagonal simultaneously after the Fourier transformation and one plane wave leads to one phonon. In order to get the CE for the amorphous solid, we need to diagonalize the matrix $F$ to get decoupled excitations. We find an unitary matrix $U$ for the diagonalizing. After that, we obtain a diagonal matrix $\Omega=U\cdot F\cdot U^{\dagger}$ and then define a new vector $P=U\cdot Q$. We multiply the both sides of Eq.(\ref{CEequQ}) to the left by the matrix $U$. And we insert the identity matrix $U^{\dagger}\cdot U$ in between the matrices $F$ and $Q$ of Eq.(\ref{CEequQ}). In this way, we get an equation
\begin{align}
\label{CEequP}
\ddot{P}=-\Omega \cdot P
\end{align}
with $\Omega$ diagonal. Now we are at the position to solve the equation (\ref{CEequP}). We take the $p$-th mode of the vector $P$ as an example. The $p$-th mode is at the $p$-th row of $P$ and is denoted by $P_p$. We also denote the entry at the $p$-th row and the $p$-th column of the diagonal matrix $\Omega$ by $\Omega_{p,p}$. Then, the equation (\ref{CEequP}) is reduced to be $\ddot{P}_p=-\Omega_{p,p}P_p$ for the mode. We set $P_p$ have a time phase of $e^{i2\pi \omega_p t}$ with $\omega_p$ as the vibration frequency for the $p$-th mode. We substitute the time phase into the reduced equation of $P_p$. We solve out that the frequency $2\pi\omega_p$ is the square root of $\Omega_{p,p}$ with a positive and real value.\\

The physical meaning of $P$ is the key element to get the CE. In the vector $Q$, each entry represents an organization of all the atoms in the real space to form a plane wave. The plane waves have various wave vectors and interfere with each other in the amorphous solid. They are not decoupled. The interference of the plane waves is reflected by the non-diagonal matrix $F$ in Eq.(\ref{CEequQ}). Then, we use the matrix $U$ to reorganize the plane waves to form wave packets in the reciprocal space. Each entry in the vector $P$ represents a wave packet reorganized by $U$. The wave packets are decoupled to each other and they are exactly the CEs we want. Each wave packet is composed of a collection of plane waves with various wave vectors but only one same vibration frequency, like $\omega_p$ in our example for the wave packet $P_p$. Physically, one CE is the collective vibration of the atoms in the real space and the atomic vibrations have various wave vectors but one vibration frequency.\\

It is clear now that a CE, say $P_p$, is a wave packet. Explicitly, we have the expression of $P_p=\sum_q U_{pq}Q_q$ for the CE basing on the definition of $P$. The contribution of each plane wave $Q_q$ to the wave packet $P_p$ is exactly revealed by the entry $U_{pq}$ which is at the $p$-th row and the $q$-th column in $U$. Generally, $U_{pq}$ is a complex number including the informations of intensity and phase of the plane wave $Q_q$.   

\subsection{Local Excitation}
Compared to the CE that is for the collective vibration of global atoms, the local excitation (LE) is localized around a single atom that stimulates the LE. The intensity of the LE decays when the LE propagates away from the center. In the amorphous solid, the disorder arrangement of atoms makes every atom be the center to stimulate LEs. Without lose of generality, we take the $0$-th atom as an example to stimulate a LE and think about the $l$-th atom which the LE can reach with $l\neq 0$. We set the equilibrium position of the $0$-th atom by $\vec{R}_0$ and that of the $l$-th atom by $\vec{R}_l$. The displacement $\vec{u}_l$ of the $l$-th atom for the LE vibration must decay when the distance $|\vec{R}_l-\vec{R}_0|$ between the $0$-th and the $l$-th atoms increases. For simplicity, we neglect the anisotropic decaying along different directions for LEs. To show the decaying, we introduce a complex wave vector $\kappa=\kappa_r+i\kappa_i$ with $\kappa_r$ and $\kappa_i$ both real scalars. The displacement for the vibration follows $\vec{u}_l=\vec{A}_{\kappa}e^{i\kappa |\vec{R}_l-\vec{R}_0|}$ with $\kappa_i$ be positive to guarantee the decaying of the LE. The coefficient $\vec{A}_{\kappa}$ can be written inversely as $\vec{A}_{\kappa}=\vec{u}_l e^{-i\kappa |\vec{R}_l-\vec{R}_0|}$. Basing on such statement, we can define a transformation.\\

Similar to the case of CE, we define the transformation to get a quantity $\mathcal{Q}_{\kappa}^{\alpha}=\sum_l \frac{1}{\sqrt{N}}e^{-i\kappa |\vec{R}_l-\vec{R}_0|}u_{l,{\alpha}}$. We multiple both sides of Eq.(\ref{dynequreduced}) by $\frac{1}{\sqrt{N}}e^{-i\kappa |\vec{R}_l-\vec{R}_0|}$ and sum over all the atoms on the both sides. After some algebra, we get a new equation from Eq.(\ref{dynequreduced}) by using the definition of $\mathcal{Q}_{\kappa}^{\alpha}$. The equation reads
\begin{align}
\label{IEequ}
\ddot{\mathcal{Q}}_{\kappa}^{\alpha}=-\sum_{\beta}\frac{1}{L_{\kappa_r'}L_{\kappa_i'}}\int \mathcal{F}_{\kappa, \kappa'}^{\alpha, \beta} \mathcal{Q}_{\kappa'}^{\beta}d \kappa'
\end{align}
with $\mathcal{F}_{\kappa, \kappa'}^{\alpha, \beta}=\sum_{l,p}e^{-i\kappa|\vec{R}_l-\vec{R}_0|}\Phi_{p,l}^{\alpha, \beta}e^{i\kappa'|\vec{R}_p-\vec{R}_0|}$. Note that $d\kappa'=d\kappa_r'd\kappa_i'$ in Eq.(\ref{IEequ}) for convenience. And, $L_{\kappa_r'}$ is the total length for $\kappa_r'$ in the reciprocal space since we have ignored the direction of the wave vectors. $L_{\kappa_i'}$ then is the total length for $\kappa_i'$. The detail derivation could be found in Appendix B.\\

In the following, we express Eq.(\ref{IEequ}) in matrix form as we have done for the CE. We arrange the components $\mathcal{Q}_{\kappa}^{\alpha}$ in a column as a vector $\mathcal{Q}$ indexed by $\kappa_r$, $\kappa_i$ and $\alpha$. To discrete the integral, we replace $d\kappa_r'$ and $d \kappa_i'$ by the infinitesimal length $\Delta \kappa_r'$ and $\Delta \kappa_i'$ respectively. And then define  $L_{\kappa_r'}/\Delta \kappa_r'=\mathcal{N}_r$ for the discrete number of the real part $\kappa_r'$ in the reciprocal space. Similarly, we set $L_{\kappa_i'}/\Delta \kappa_i'=\mathcal{N}_i$ for the imaginary part $\kappa_i'$. Note that $\kappa_i'$ must be positive while $\kappa_r'$ can be both of positive and negative. Finally, we define a matrix $\mathcal{F}$ for $\mathcal{F}_{\kappa, \kappa'}^{\alpha, \beta}$ by absorbing the factor $1/(\mathcal{N}_r\mathcal{N}_i)$. Then, we have an equation transformed from Eq.(\ref{IEequ}), reading
\begin{align}
\label{LEequQ}
\ddot{\mathcal{Q}}=-\mathcal{F}\cdot \mathcal{Q}.
\end{align}
Similar to the matrix $F$ for CE, local vibrations with various wave vectors in $\mathcal{F}$ are interfered with each other. We need diagonalize $\mathcal{F}$ to get decoupled LEs. We find an unitary matrix $\mathcal{U}$ for the diagonalizing and then get a diagonal matrix $\Lambda=\mathcal{U}\cdot \mathcal{F} \cdot \mathcal{U}^{\dagger}$. We define a vector by $\mathcal{P}=\mathcal{U} \cdot \mathcal{Q}$ for the LE. Finally, we have an equation for the LE, reading
\begin{align}
\label{IEequP}
\ddot{\mathcal{P}}=-\Lambda \cdot \mathcal{P}.
\end{align}
To solve this equation, we still take the time phase $e^{i2\pi\omega_p t}$ for the entry $\mathcal{P}_p$ at the $p$-th row in $\mathcal{P}$ with $\omega_p$ the vibration frequency. Then from Eq.(\ref{IEequP}), $2\pi \omega_p$ is the square root of $\Lambda_{pp}$ that is at the $p$-th row and $p$-th column in $\Lambda$. \\

Each entry in $\mathcal{P}$ represents an LE. Each LE comprises a collection of decaying plane waves. Those decaying plane waves for the LE are with various complex wave vectors but only one vibration frequency. The imaginary part of the complex wave vectors decays the LE away from the stimulating center. The contributions of the decaying plane waves to an LE are calculated from the matrix $\mathcal{U}$, as we have specified for the CE by using the matrix $U$. \\

\section{computational details}
We study CEs and LEs in an amorphous silicon as an application of our theory. The amorphous silicon is generated by MD. The Eigen library is implemented in our code to diagonalize matrices. In the following, we specify the computational details.
\subsection{Amorphous Silicon}
We use MD to generate an amorphous silicon. The MD simulations were performed using the Large-scale Atomic/Molecular Massively Parallel Simulator (LAMMPS) with a time step of 0.5 $fs$~\cite{lmps1,lmps2}. In the simulations, the Stilling-Web interatomic potential was implemented and the periodical boundary condition was applied~\cite{sw}. We follow the simulation steps in Ref.(\onlinecite{moon2}) for LAMMPS . We started from a crystalline silicon with 8 cells along $[001]$, $[010]$ and $[100]$ directions, containing 4096 silicon atoms in total, and melt the structure at 3500 $K$ for 500 $ps$ in an NVT ensemble. Next, we quenched the liquid silicon to 1000 $K$ with the quench rate of 100 $K/ps$, followed by annealing the structure at 1000 $K$ for 25 $ns$. Finally, we quenched the structure at a rate of 100 $K/ps$ to 300 $K$ and then equilibrated the structure at 300 $K$ for 10 $ns$ in an NVT ensemble. After an additional equilibration at $300K$ in an NVE ensemble for 500 $ps$, we obtain the amorphous silicon we need.\\

\subsection{Dynamics Matrix}
We consider the force constant $\Phi_{l,p}^{\alpha, \beta}=(\frac{\partial^2 \Psi}{\partial R_{l,\alpha}\partial R_{p,\beta}})(\frac{1}{\sqrt{M_lM_p}})$ for the $l$-th atom vibrating along $\alpha$ direction and the $p$-th atom along $\beta$ direction with $l\neq p$. We fix the $l$-th atom at some position that is shift away from its equilibrium position along $\alpha$ direction. And then shift the $p$-th atom away from its equilibrium position along $\beta$ direction. In shifting the $p$-th atom, we calculate the difference of the total interatomic potential $\Psi$ at various positions of the $p$-th atom to get the derivative $\partial \Psi/\partial R_{p,\beta}$. Then, we fix the $l$-th atom at a new position along $\alpha$ direction and repeat shifting the $p$-th atom to get a new derivative of $\partial \Psi/\partial R_{p,\beta}$. Basing on the difference of the derivative $\partial \Psi/\partial R_{p,\beta}$ for the $l$-th atom at various positions, we get the second order of the derivative $(\frac{\partial^2 \Psi}{\partial R_{l,\alpha}\partial R_{p,\beta}})$ for the force constant. In the calculation, the Stilling-Web interatomic potential is applied~\cite{sw}. For the case of $l=p$, the force constant $\Phi_{l,l}^{\alpha, \beta}=-\sum_{p\neq l}\Phi_{l,p}^{\alpha, \beta}$ has been well defined to guarantee that no force is applied on the $l$-th atom along $\alpha$ direction when all the atoms move by an identical distance along $\beta$ direction. Based on the above statement, the dynamics matrix is obtained.\\

In diagonalizing matrices $F$ and $\mathcal{F}$, the unitary matrices $U$ and $\mathcal{U}$ obtained are complex, including the informations of intensity and phase of waves for the excitations. The frequencies of the excitations must be real and positive as well. Thus, after the diagonalizing, we only take the real and positive entries from the matrices $\Omega$ and $\Lambda$ for the frequencies of the excitations. In the numerical calculation, we set a small value $\epsilon$ and select the entries with their imaginary parts in the range of $(-\epsilon, \epsilon)$ from the matrices $\Omega$ and $\Lambda$.  \\

\subsection{Structure of Excitations}
\label{stru}
We focus on the physical structure of the excitations. We take the $p$-th CE as an example, which is the $p$-th entry of the vector $P$ and has the expression of $P_p=\sum_q U_{pq}Q_q$. As we have discussed, $U_{pq}$ is the contribution of the plane wave $Q_q$ to the CE of $P_p$. $U_{pq}$ is a complex number, have the informations of intensity and phase of the wave $Q_q$. In this study, we take the absolute value of $U_{pq}$ to show the intensity of the wave $Q_q$, and neglect the phase information. By solving Eq.(\ref{CEequP}), we get the frequency $\omega_p$ for the CE of $P_p$. We note the wave vector of $Q_q$ by $\kappa_q$. By manipulating the relation of $\omega_p$, $\kappa_q$ and the intensity $|U_{pq}|$ of $Q_q$, we can investigate the structure of the CE $P_p$ in the reciprocal space. Such statement can be applied on LE with the same treatment.\\

For a given wave vector, say along $x$ direction, there exists two types of CEs. One type of the CE is transverse, in which atoms have the vibrational direction perpendicular to the wave vector. The transverse CE has two degenerate states since there exist two orthogonal directions $y$ and $z$ both normal to $x$. The other type CE is longitudinal, in which atoms have the vibrational direction parallel to the wave vector. To illustrate our theory, we vary the wave vector along only one direction such as only along $x$ direction. In this case, the CEs composed of $Q_{\vec{\kappa}}^y$ (or $Q_{\vec{\kappa}}^z$) is the transverse CEs and the CEs composed of $Q_{\vec{\kappa}}^x$ are for the longitudinal CEs.\\

In this calculation,we normalize the mass of one silicon atom by $10^{-26}kg$, and the length by $1\AA$. The energy is normalized by $10KJ/mol$, which is $1.66\times 10^{-20}J$ per atom. Thus, the wave vector is normalized by $\AA^{-1}$ and the frequency for the normalization is $2.1 THz$.

\section{results}
\label{res}
\begin{figure}[!h]
\includegraphics[width=0.4\textwidth]{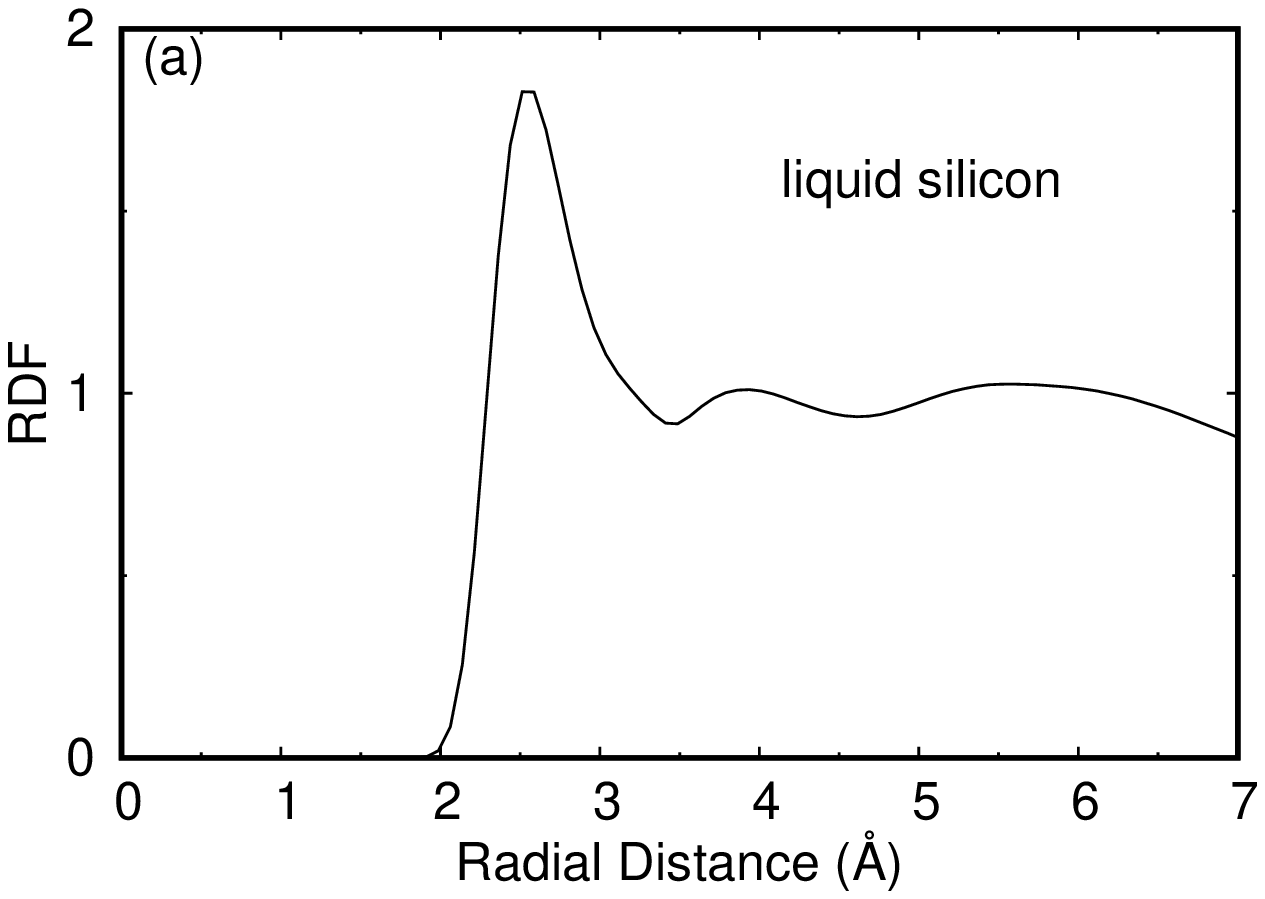}\\[0.5cm]
\includegraphics[width=0.4\textwidth]{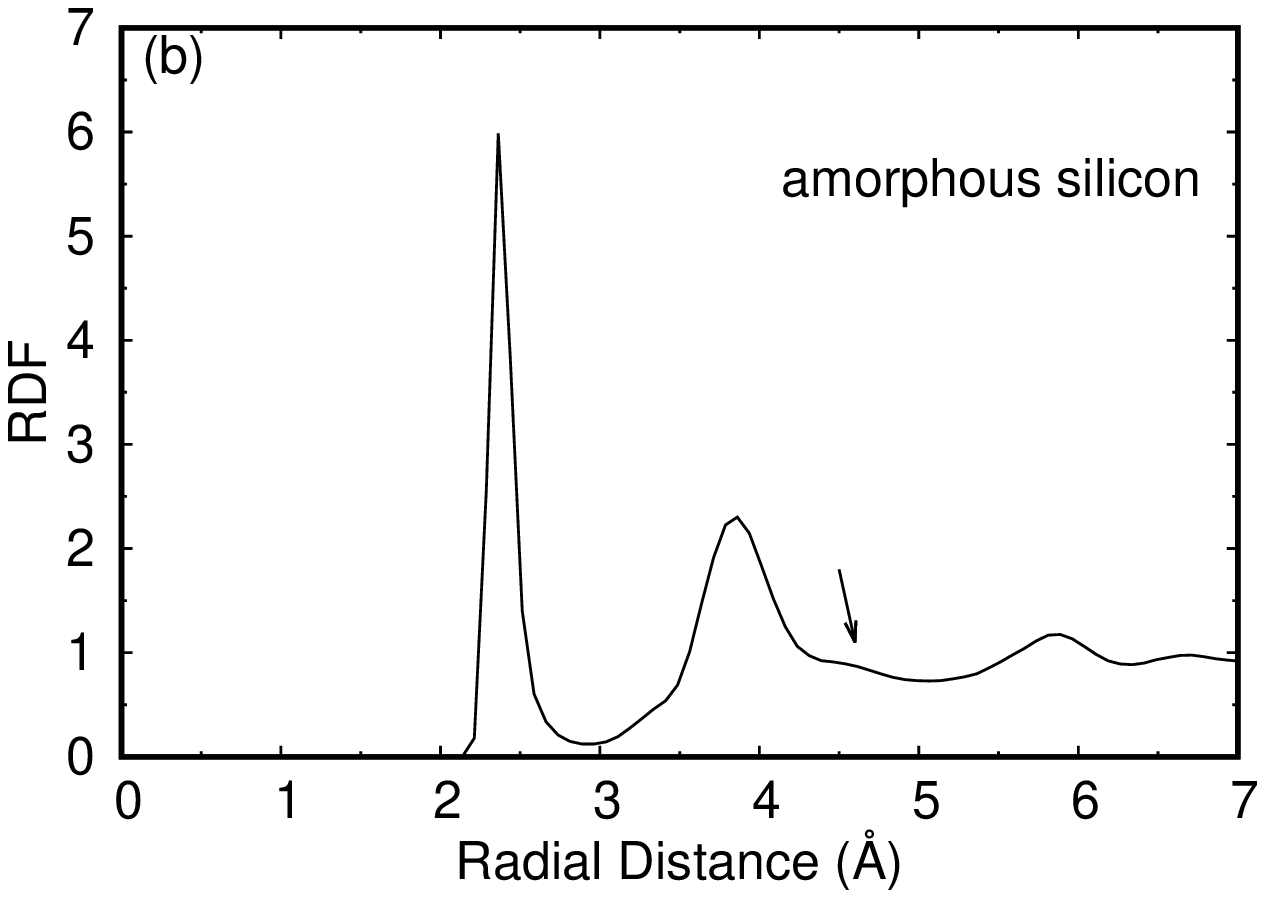}
\caption{\label{rdfofAS} Radial Distribution Function (RDF) of silicon. (a) Liquid silicon is melted at $3500K$. (b) Amorphous silicon is obtained at $300K$ by LAMMPS. The split of the second peak in (b) has been indicated by an arrow.}
\end{figure}

We use LAMMPS to generate the amorphous silicon. Fig.(\ref{rdfofAS}a) is the Radial Distribution Function (RDF) of the liquid silicon melted at $3500K$ as the first step in the LAMMPS simulation. The RDF of the amorphous silicon at $300K$ is plotted in Fig.(\ref{rdfofAS}b), in which the first peak is located at $2.3\AA$ and shows the local order of the structure. The second peak of the RDF in Fig.(\ref{rdfofAS}b) is split, which is the main feature of the amorphous solid different from the liquid RDF in Fig.(\ref{rdfofAS}a)~\cite{zal}. The split of the second peak has been indicated by an arrow in Fig.(\ref{rdfofAS}b) for clarity, meaning that the silicon we study is really an amorphous solid.

\subsection{Collective Excitation}
As we have mentioned in Section(\ref{stru}), the solutions to Eq.(\ref{CEequP}) give us the following informations for the $p$-th mode CE, the frequency $\omega_p$, the wave vector $\kappa_q$ for the wave $Q_{q}$ contributing to the CE, and the intensity $|U_{pq}|$ of $Q_{q}$. In the following plots, we drop off the subscripts $p$ and $q$ for clear notation. And we use the phrase of intensity $|U|$ referred to the intensity $|U_{pq}|$ of $Q_{q}$ in short. We manipulate the relation of $\omega$, $\kappa$ and intensity $|U|$ to show the structures of CEs. 
\begin{figure}[!h]
\includegraphics[width=0.5\textwidth]{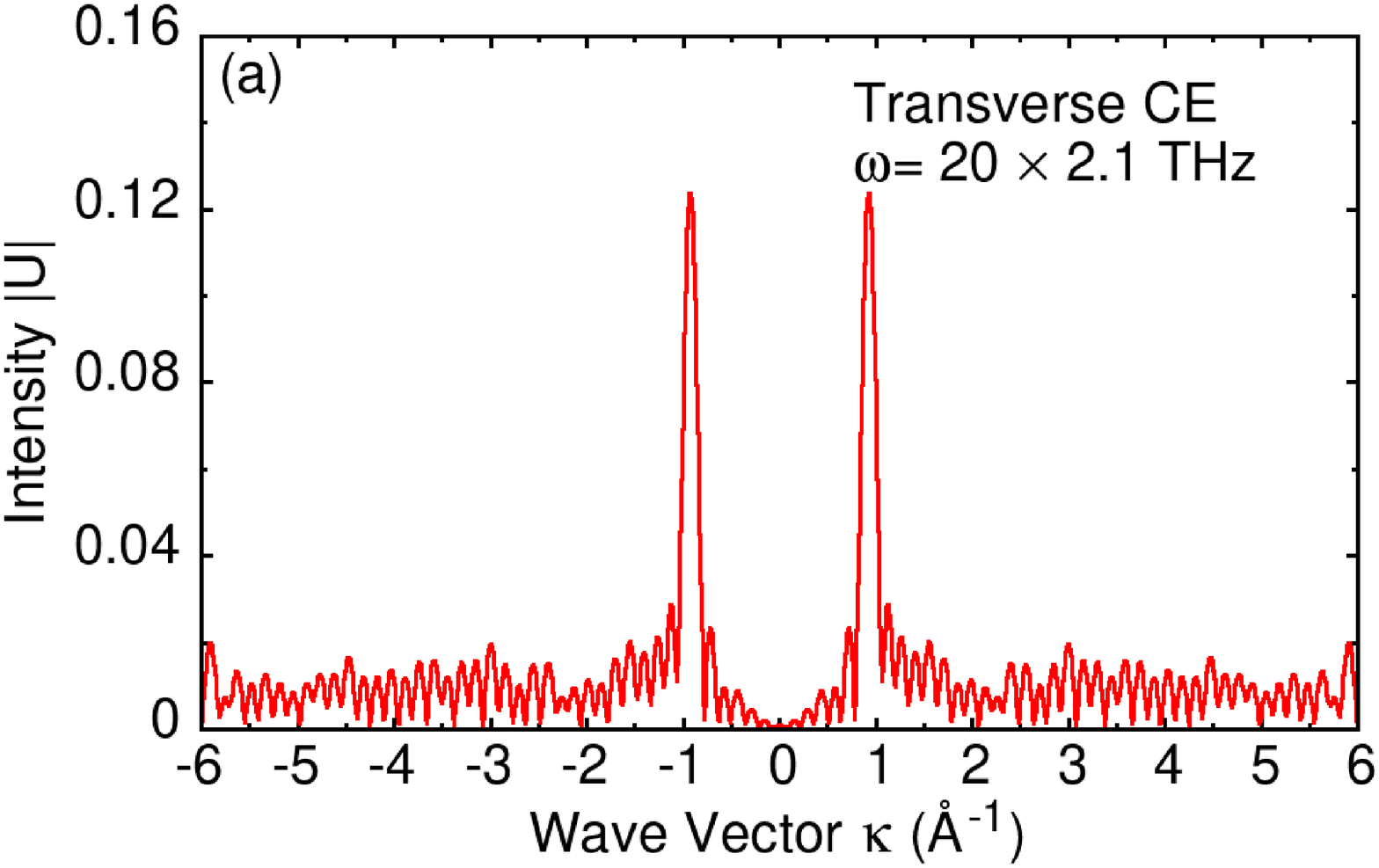}
\includegraphics[width=0.5\textwidth]{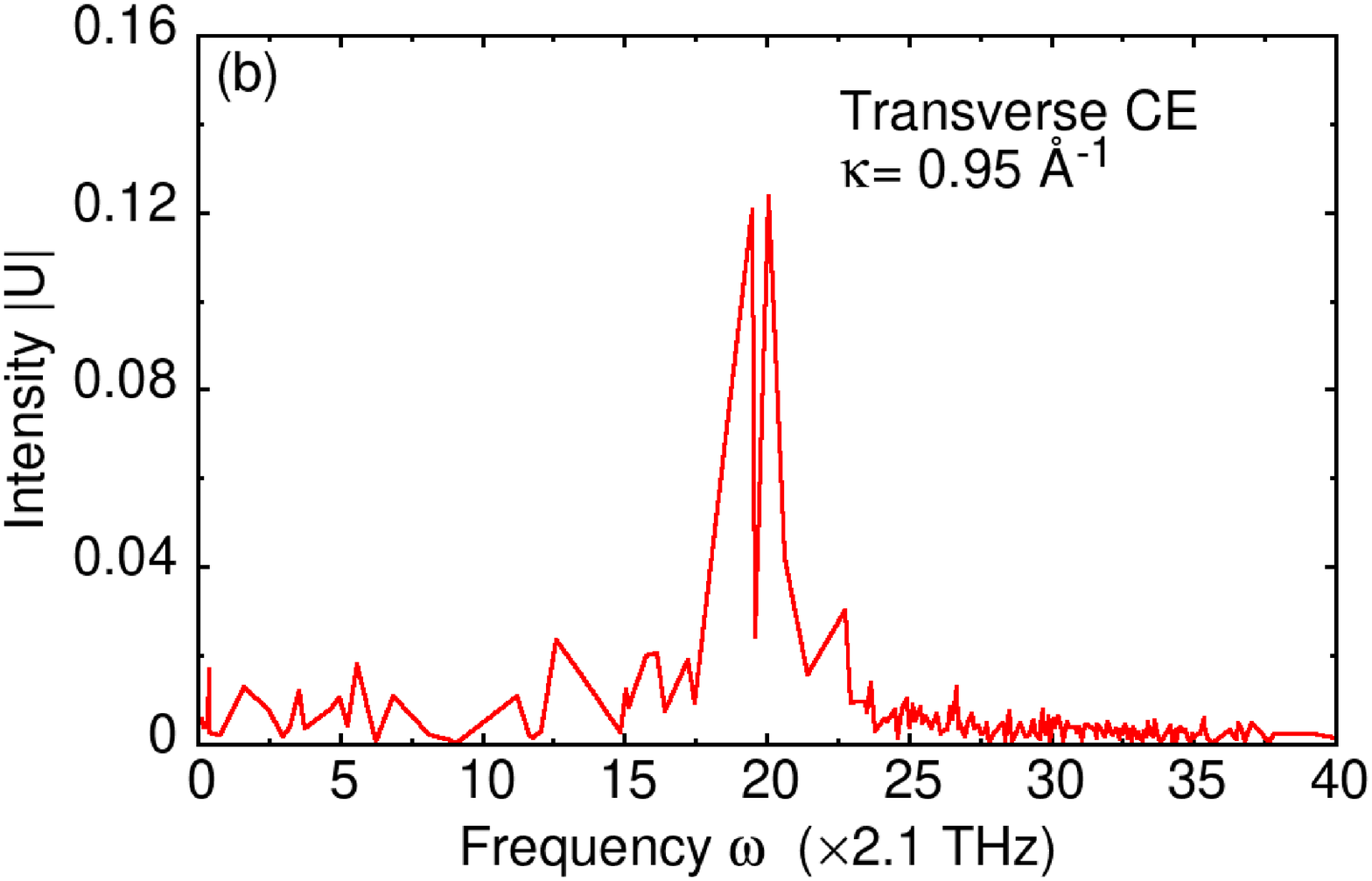}
\caption{\label{tansPE} Intensity $|U|$ of transverse CEs. (a) For an transverse CE with the frequency fixed at $20\times 2.1 THz$, the intensity of the CE shows two wave packets in the reciprocal space. (b) For transverse CEs with the wave vector $\kappa$ fixed at $0.95 \AA^{-1}$, the intensity of the CEs shows a wave packet by scanning the frequency. }
\end{figure}
In Fig.(\ref{tansPE}a), we fix the frequency $\omega=20\times 2.1 THz$ for a transverse CE and plot intensity $|U|$ as functional of the wave vector $\kappa$. Here $2.1 THz$ is the frequency normalization as we have mentioned in Section(\ref{stru}). In the figure, we get two wave packets and the peaks of the two wave packets are located at $-0.95 \AA^{-1}$ and $0.95 \AA^{-1}$ respectively. These two wave packets are symmetric about $\kappa=0$ and they are the forward and the backward waves respectively. Due to the symmetry, we focus on only one wave packet at the peak of $0.95 \AA^{-1}$. The width of the half intensity of the peak is about $0.2 \AA^{-1}$, which is corresponding to a wave packet with a scale of $2\pi/0.2 \approxeq 31\AA$ in the real space. Then, we fix $\kappa=0.95 \AA^{-1}$ in the reciprocal space and plot the intensity  $|U|$ in Fig.(\ref{tansPE}b) by scanning the frequency $\omega$. We still get a peak, meaning that for a well defined wave vector $\kappa$ there exist many waves $Q$ that have various vibration frequencies. Comparably, in crystalline solids, an excitation of lattice vibration or a phonon has only one well defined wave vector corresponding to one frequency. Such difference of the excitation structure of the atomic vibrations in a amorphous solid and a crystalline solid is due to the disorder arrangement of atoms in the former. \\

\begin{figure}[!h]
\includegraphics[width=0.5\textwidth]{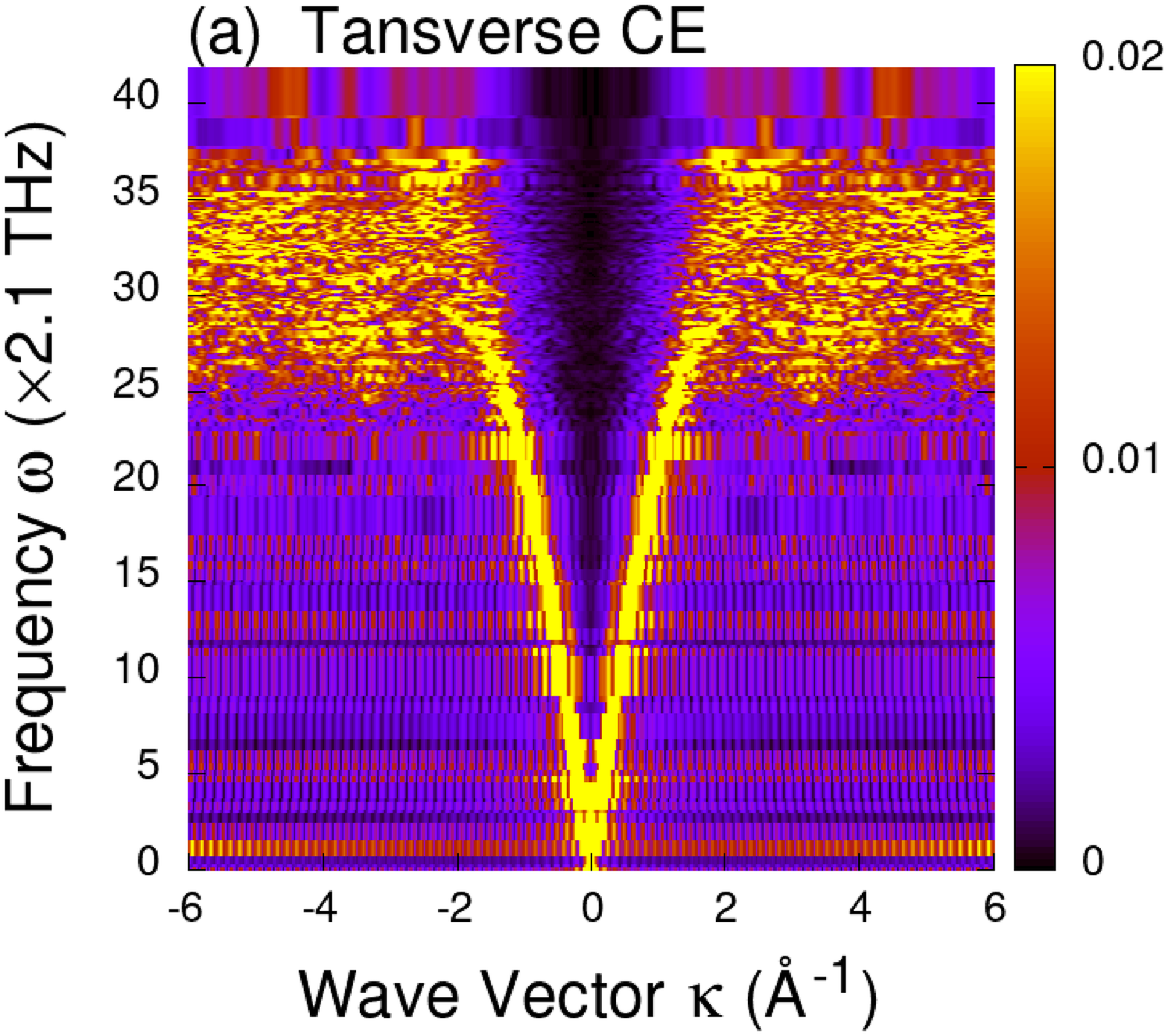}
\includegraphics[width=0.5\textwidth]{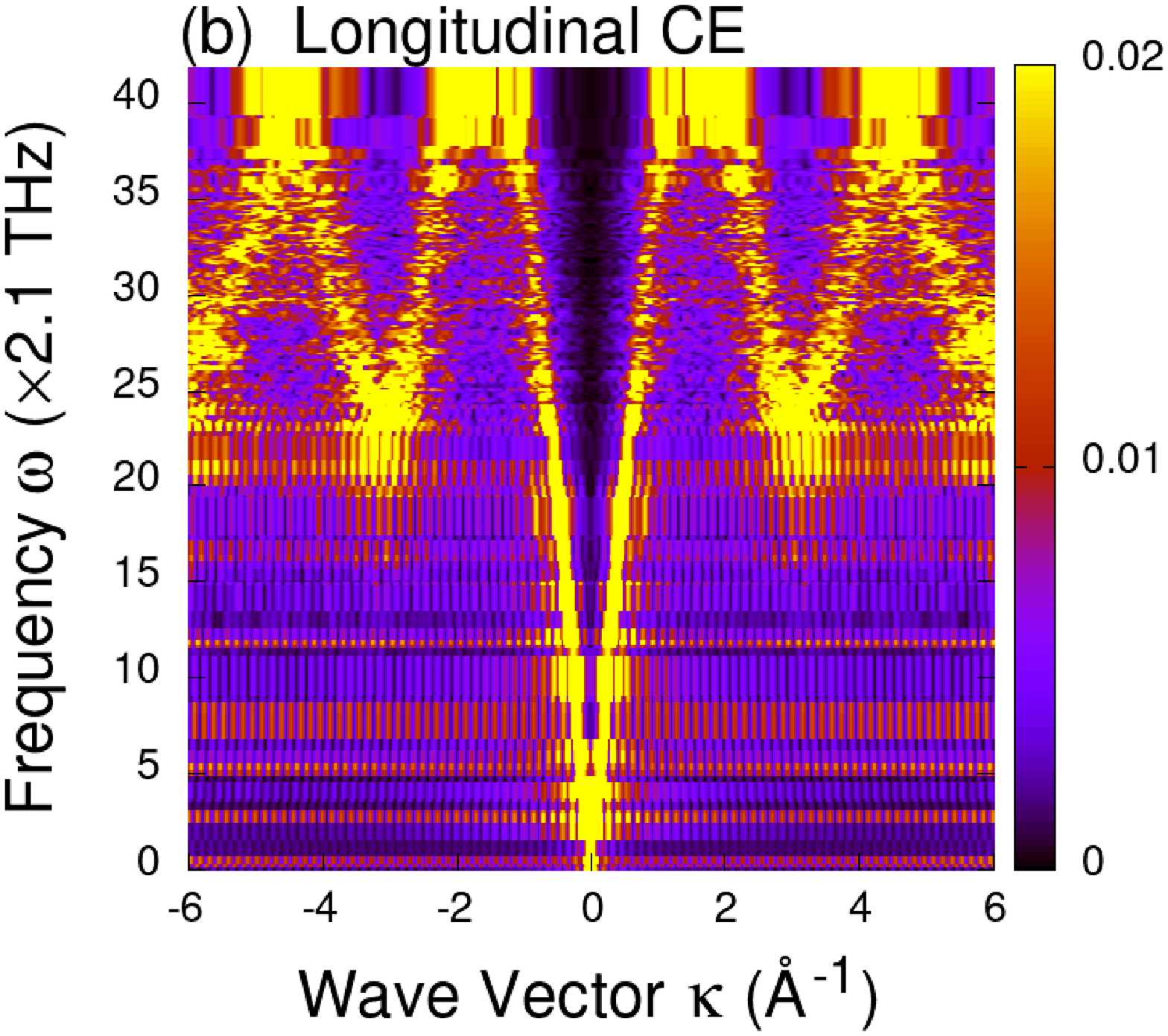}
\caption{\label{dispPE} Structure of collective CEs in the reciprocal space. (a) Structure of the transverse CEs in the reciprocal space shows two branches starting from zero frequency to $\omega=25\times 2.1 THz$ and end into a frequency band. (b) A periodical structure is found for the longitudinal CEs in the reciprocal space.}
\end{figure}

We use a color bar to show the intensity $|U|$ of the CEs and plot the relation of the frequency $\omega$ and the wave vector $\kappa$ in Fig.(\ref{dispPE}). Fig.(\ref{dispPE}a) is for the transverse CE and Fig.(\ref{dispPE}b) is for the longitudinal CE. 
Results show that the frequency $\omega$ of CEs is no more than $42\times 2.1 THz$. In Fig.(\ref{dispPE}a), we observe two branches start from $\omega=0 THz$ to $\omega=25\times 2.1 THz$, behaving like the acoustic branches in the crystalline silicon. The branches in Fig.(\ref{dispPE}a) are linear-like due to the disorder arrangement of atoms erasing the anisotropic scattering of waves in the amorphous silicon. Each branch has a broadening line width, which shows that the CEs are wave packets as we have given the example in Fig.(\ref{tansPE}). Once the frequency is larger than $\omega=25\times 2.1 THz$, the two branches end into a band in Fig.(\ref{dispPE}a). The upper frequency of the band is $\omega=37\times 2.1 THz$. In the band, the structure of the transverse CEs is random rather than a wave packet. This is because the transverse CEs in the band have large wave vectors and short wave lengths. Those transverse CEs with short wave lengths can see the discreteness of the atoms in the solid and can be scattered by the atoms easily. What is more, those transverse CEs vibrate with the directions perpendicular to the wave vector. The random arrangement of atoms along the vibration directions of the transverse CEs multi-scatters the CEs for the propagation along the wave vector, which brings the random structure of the transverse CEs in the reciprocal space. On the other side, the occurrence of the branches in Fig.(\ref{dispPE}a) is due to the long wave length of the transverse CEs by which the CEs can not distinguish the discreteness of the atoms. It also could be found in the figure that there exists a gap between the two branches. That means no transverse CEs can be stimulated in the gap.\\

It is very interesting to find a periodical structure in Fig.(\ref{dispPE}b) for longitudinal CEs. In the figure, we still can find two branches go up from $\omega=0 THz$, but end at $\omega=42\times 2.1THz$. And then the two branches go down with a zigzag structure. The zigzag structure is periodical with the period roughly about $2.8\AA^{-1}$ since we can not accurately locate the peaks for such disorder system. We think the phase for one period is $2\pi$, and calculate the averaged lattice period in the real space corresponding to the periodicity in the reciprocal space. We find that the averaged lattice period is about $2.3 \AA$, which is the location of the first peak in Fig.(\ref{rdfofAS}b). Such observation reveals that the local order of the structure in the amorphous solid plays its role as the lattice parameter and makes the amorphous solid behave like a crystalline solid for the longitudinal CEs. Thus, it is possible for us to define a quasi-Brillouin zone for the longitudinal CEs. The boundary of the first quasi-Brillouin zone is at $\pm \pi/a$ with $a$ the location of the first peak in RDF. Then, we can go further to define the reciprocal lattice vector by $G=2\pi/a$ for the longitudinal CEs, and map the longitudinal CEs to the phonons of crystalline solids for the study of physical properties. Such work is out of the scope of this paper.\\      

\subsection{Local Excitation}
\begin{figure}[!t]
\includegraphics[width=0.4\textwidth]{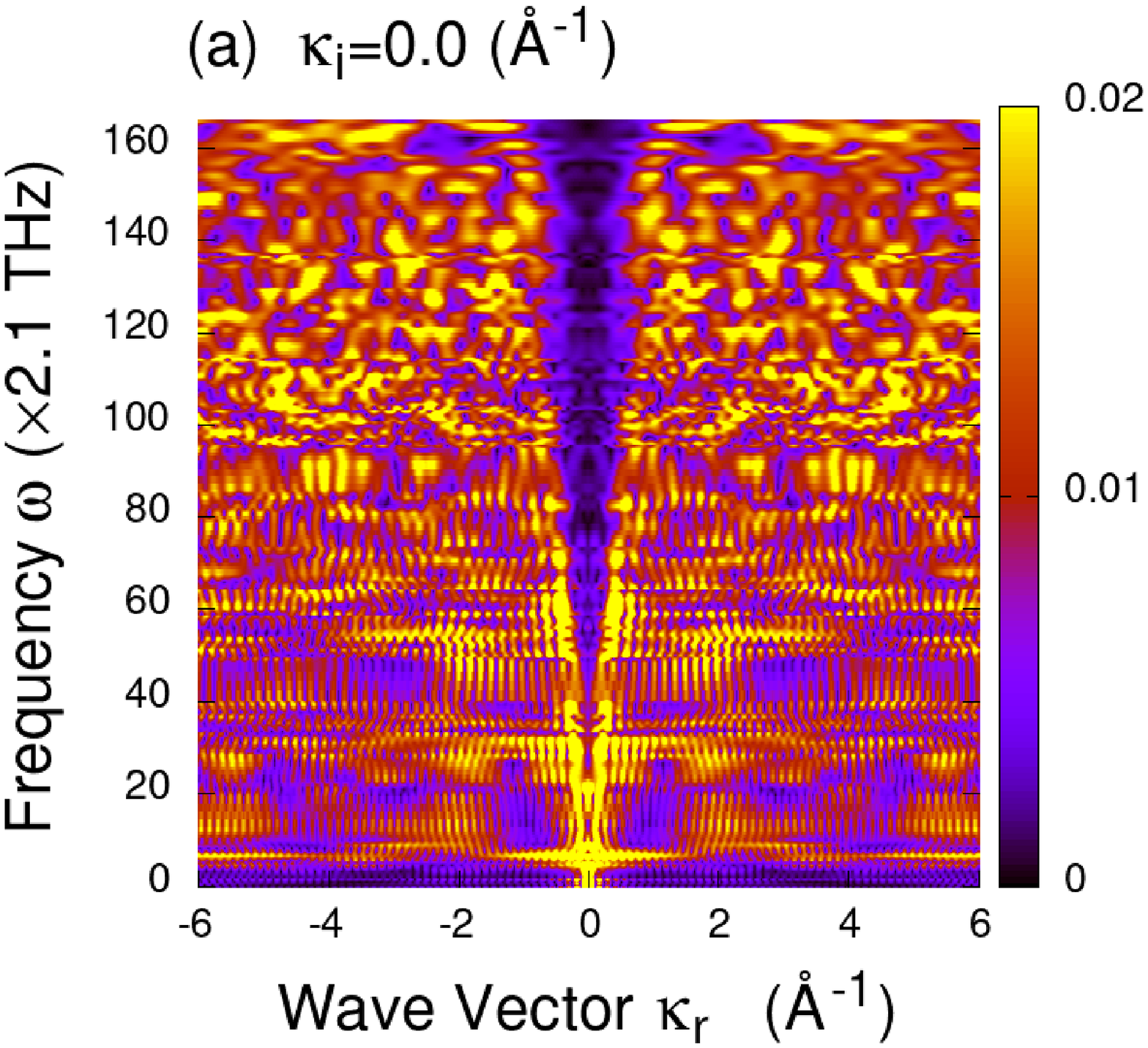}
\includegraphics[width=0.4\textwidth]{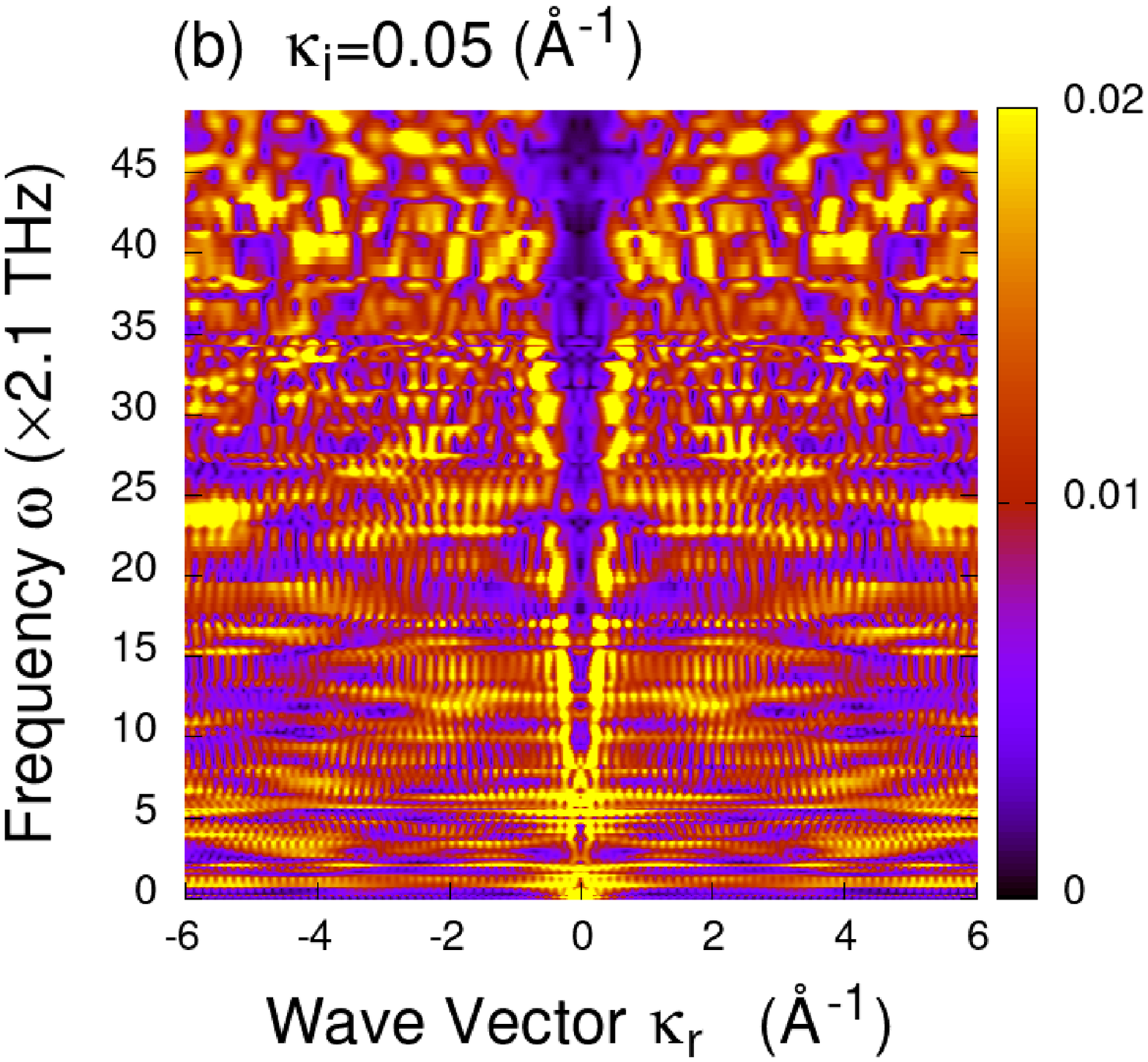}
\includegraphics[width=0.4\textwidth]{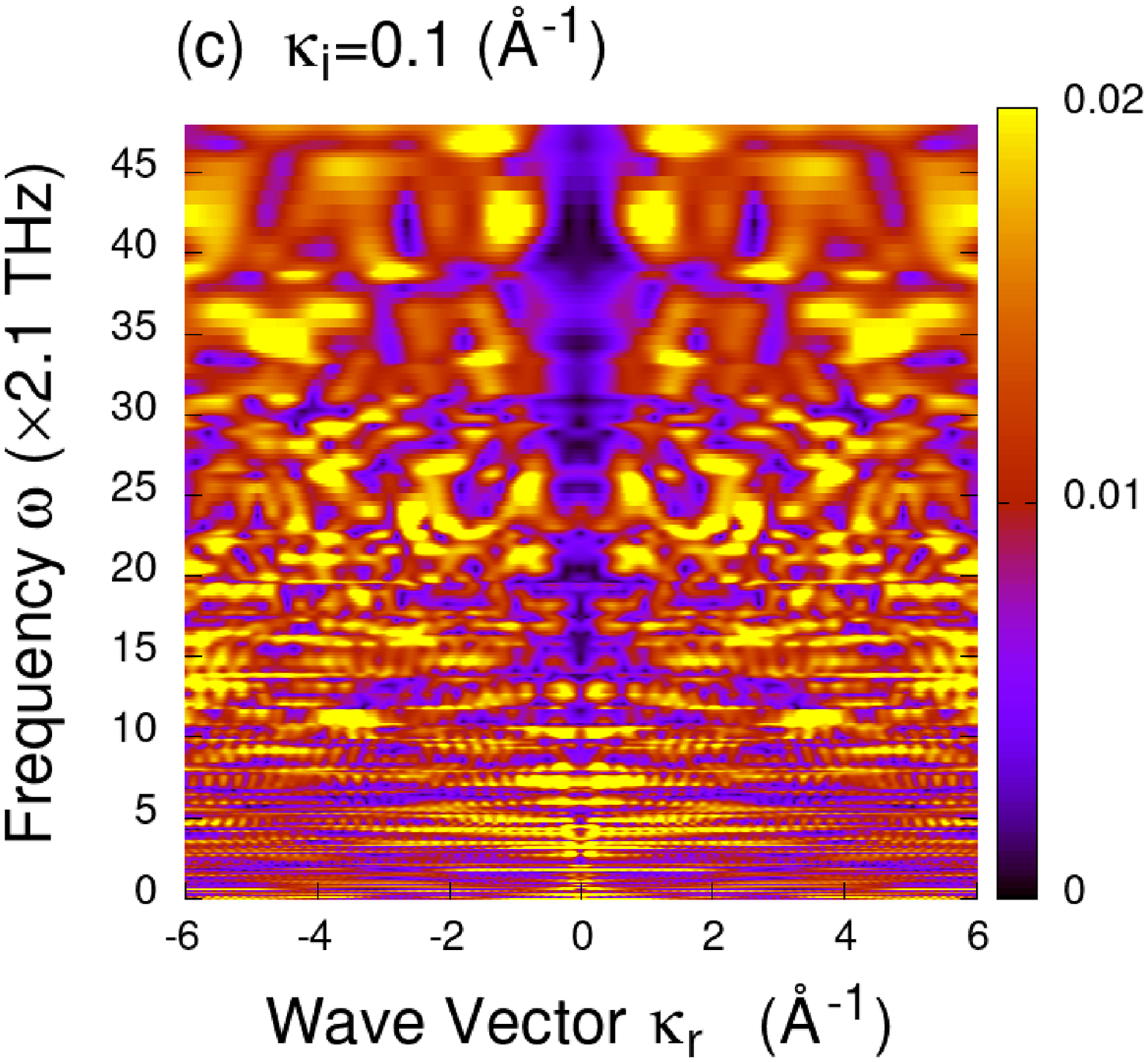}
\includegraphics[width=0.4\textwidth]{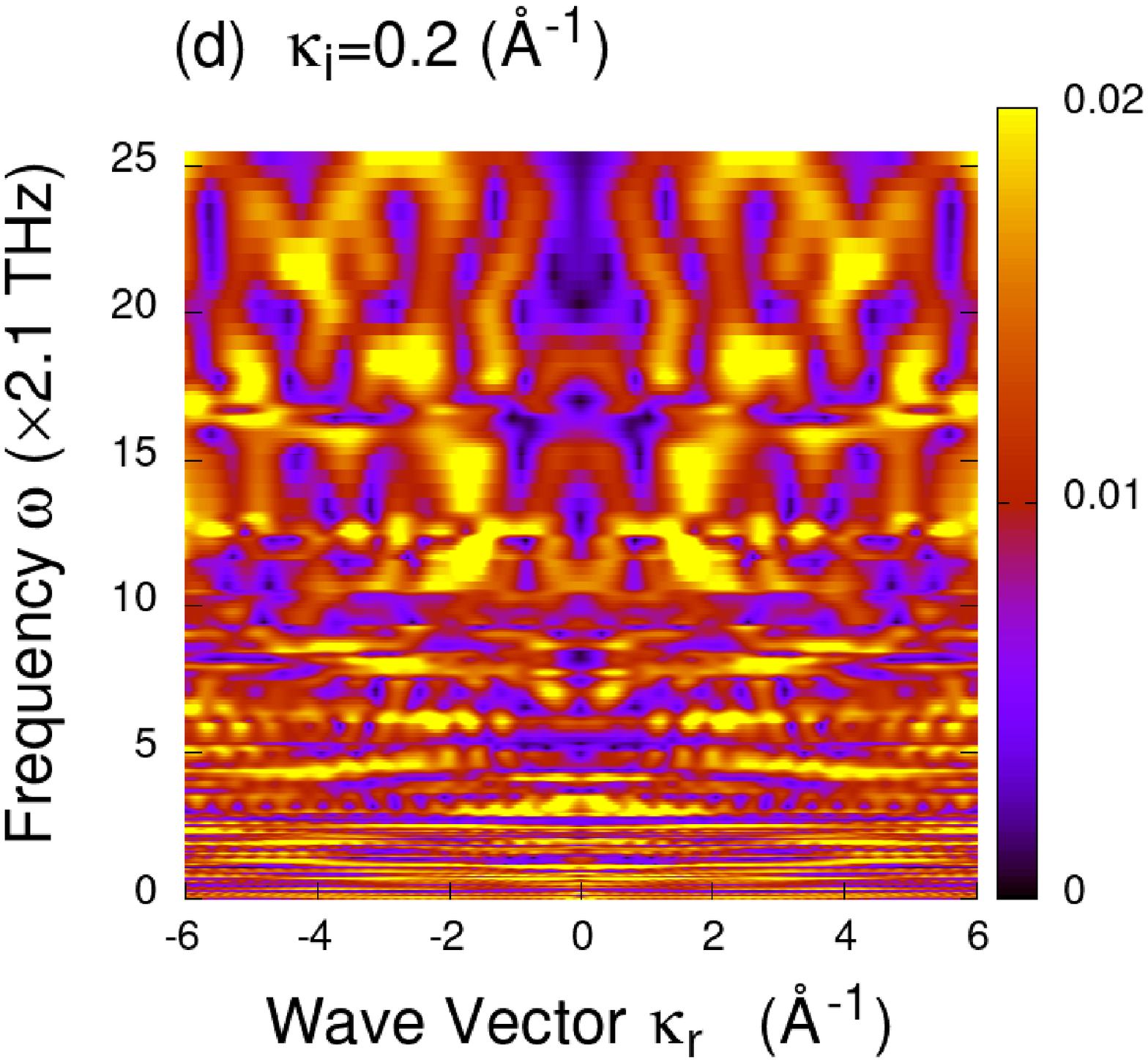}
\caption{\label{dispLE} Structure of LEs in the reciprocal space. The imaginary part $\kappa_i$ of the wave vector is fixed for each map. (a) $\kappa_i=0.0 \AA^{-1}$. (b) $\kappa_i=0.05 \AA^{-1}$. (c)$\kappa_i=0.1 \AA^{-1}$. (d)$\kappa_i=0.2 \AA^{-1}$.}
\end{figure}

To reveal the structures of LEs, we need to show the relation of the frequency $\omega$, the wave vector $\kappa$, and the intensity $|\mathcal{U}|$. Note that $\kappa$ of LEs is a complex number with the real part $\kappa_r$ and the imaginary part $\kappa_i$. That means we have four parameters at hand for the study. To clearly show the structure of the LEs, we use the color bar to show the intensity $|\mathcal{U}|$ and fix $\kappa_i$ for each map. We show the results in Fig.(\ref{dispLE}). 
In Fig.(\ref{dispLE}a), we show the structure of LE with $\kappa_i=0\AA^{-1}$. Zero $\kappa_i$ means the LE stimulated by a given atom can propagate without decaying. In the figure, the frequency for the LEs can reach as high as $170\times 2.1 THz$, which is much higher than the maximum frequency of CEs. Those LEs stimulated locally by the atoms are distributed randomly in the reciprocal space as shown in Fig.(\ref{dispLE}a). The LEs interfere with each other in the amorphous solid to form CEs eventually as we have shown in Fig.(\ref{dispPE}). We also find a gap close to $\kappa_r=0$ in Fig.(\ref{dispLE}a), which originates the gap of the CEs in Fig.(\ref{dispPE}).\\

In the case of $\kappa_i\neq 0 $, LEs will decay their intensities in propagating. The decaying length is approximated to be $1/\kappa_i$. Such as in Fig.(\ref{dispLE}b), we have $\kappa_i=0.05\AA^{-1}$ and the decaying length is about $20\AA$. It could be found that the maximum frequency in Fig.(\ref{dispLE}b) is $69\times 2.1 THz$ that is much lower than the maximum frequency in Fig.(\ref{dispLE}a). The LEs with frequencies higher than $69\times 2.1 THz$ do not satisfy Eq.(\ref{IEequP}) when $\kappa_i=0.05\AA^{-1}$. This is because the LEs with a higher frequency have a larger energy and can propagate further out of the decaying length of $20\AA$ under the condition of $\kappa_i=0.05\AA^{-1}$. The LEs are still randomly distributed in the reciprocal space and the gap still can be found in Fig.(\ref{dispLE}b). When $\kappa_i$ is increased to a larger value, the maximum frequency for LEs decreases to a smaller value, as shown in Fig.(\ref{dispLE}c) and Fig.(\ref{dispLE}d).

\section{conclusions}  
We have studied the excitations of atomic vibrations in the reciprocal space for amorphous solids. The excitations can be classified into two categories, collective excitation and local excitation. The collective excitation is due to the collective vibrations of all the atoms in the amorphous solids while the local excitation is stimulated by a single atom locally. The wave vector for the collective excitations must be real while the wave vector for the local excitations is complex. The imaginary part of the wave vector for the local excitations decays the excitations.\\

The excitations are wave packets in amorphous solids, comprising a collection of plane waves with various wave vectors but one vibration frequency. The collective excitation has two types, the transverse excitation and the longitudinal excitation. It is interesting to find that the longitudinal excitation has a periodical structure in the reciprocal space. The periodicity is originated from the local order of the structure in the real space. The local excitation can also be found by our theory in the reciprocal space. Results show that the local excitations with higher frequencies have larger decaying lengths. For the excitations, a gap can be found in the reciprocal space where no excitation can occur.\\

In this study, we didn't touch two problems. The first problem is how to use the excitations in our theory to classify the propagons, diffusons and locons. To define the density of state of the excitations is the key to the problem. The second problem is how to calculate the thermal conductivity of the amorphous solids by using the excitations. To solve this problem, we need to apply the Bose-Einstein statistics correctly for the excitations. To solve the two problems are our future works. \\ 

The author kindly acknowledges Prof. Ning-Hua Tong from Renmin University of China for discussions.\\    
\appendix
\section{}
We multiple the both sides of Eq.(\ref{dynequreduced}) by $(1/\sqrt{N})e^{-i\vec{\kappa}\cdot \vec{R}_l}$ and sum the both sides over the total atoms. Then we get a new equation reading
\begin{align}
\label{appA1}
\ddot{Q}_{\vec{\kappa}}^{\alpha}=-\sum_{l}\frac{1}{\sqrt{N}}e^{-i\vec{\kappa}\cdot \vec{R}_l}\sum_{p,\beta}\Phi_{l,p}^{\alpha, \beta} \sum_{m}\delta_{m,p}u_{m,\beta}
\end{align}
The left hand side of Eq.(\ref{appA1}) is $Q_{\vec{\kappa}}^{\alpha}$ as we have defined in the text of this paper. On the right hand side of Eq.(\ref{appA1}), we have introduced the Kronecker Delta function $\delta_{m,p}$ to replace $u_{p,\beta}$ by $\sum_{m}\delta_{m,p}u_{m,\beta}$. The function $\delta_{m,p}$ can be expressed as
\begin{align}
\label{appA2}
\delta_{m,p}=\frac{1}{V_{\vec{\kappa}'}}\int e^{i\vec{\kappa}'\cdot (\vec{R}_p-\vec{R}_m)}d \vec{\kappa}'.
\end{align}
The integration is over the volume $V_{\vec{\kappa}'}$ in the reciprocal space. We substitute Eq.(\ref{appA2}) into Eq.(\ref{appA1}) and rewrite Eq.(\ref{appA1}) as
\begin{widetext}
\begin{align}
\ddot{Q}_{\vec{\kappa}}^{\alpha}&=-\sum_{l,p,m,\beta}\frac{1}{\sqrt{N}}e^{-i\vec{\kappa}\cdot \vec{R}_l}\Phi_{l,p}^{\alpha, \beta} \frac{1}{V_{\vec{\kappa}'}}\int e^{i\vec{\kappa}'\cdot (\vec{R}_p-\vec{R}_m)}d \vec{\kappa}'u_{m,\beta}\nonumber \\
&=-\sum_{\beta}\frac{1}{V_{\vec{\kappa}'}}\int \left [ \sum_{l,p}e^{-i\vec{\kappa}\cdot \vec{R}_l}\Phi_{l,p}^{\alpha, \beta}e^{i\vec{\kappa}'\cdot \vec{R}_p}\right ]\left [\frac{1}{\sqrt{N}}\sum_m e^{-i\vec{\kappa}'\cdot \vec{R}_m} u_{m,\beta}\right ]d\vec{\kappa}'.
\end{align}
\end{widetext}
We define $F_{\vec{\kappa}, \vec{\kappa}'}^{\alpha,\beta}=\sum_{l,p}e^{-i\vec{\kappa}\cdot \vec{R}_l}\Phi_{l,p}^{\alpha, \beta}e^{i\vec{\kappa}'\cdot \vec{R}_p}$ for the first bracket and replace the term in the second bracket by $Q_{\vec{\kappa}'}^{\beta}=\frac{1}{\sqrt{N}}\sum_m e^{-i\vec{\kappa}'\cdot \vec{R}_m} u_{m,\beta}$ as we have defined. Then, we recover Eq.(\ref{CEequ}).

\section{}
We multiple both sides of Eq.(\ref{dynequreduced}) by $\frac{1}{\sqrt{N}}e^{-i\kappa |\vec{R}_l-\vec{R}_0|}$ and sum the both sides over all the atoms. Then, we have an equation, reading
\begin{align}
\label{ldab1}
\ddot{\mathcal{Q}}_{\kappa}^{\alpha}=-\sum_l \frac{1}{\sqrt{N}}e^{-i\kappa |\vec{R}_l-\vec{R}_0|}\sum_{p,\beta}\Phi_{l,p}^{\alpha,\beta}\sum_{m}\delta_{m,p}u_{m,\beta}.
\end{align}
Here, we have introduced the Kronecker Delta function
\begin{widetext}
\begin{align}
\label{delt}
\delta_{m,p}=\frac{1}{L_{{\kappa'_r}}L_{{\kappa'_i}}}\int e^{-i\kappa'_r(|\vec{R}_m-\vec{R}_0|-|\vec{R}_p-\vec{R}_0|)}e^{\kappa'_i(|\vec{R}_m-\vec{R}_0|-|\vec{R}_p-\vec{R}_0|)}d\kappa'_rd\kappa'_i=\frac{1}{L_{\kappa'_r}L_{\kappa'_i}}\int e^{-i\kappa'(|\vec{R}_m-\vec{R}_0|-|\vec{R}_p-\vec{R}_0|)}d \kappa'.
\end{align}
\end{widetext}
Here,$\kappa'_r$ is the real part of $\kappa'$ while $\kappa'_i$ is the imaginary part of $\kappa'$. We use $d \kappa'$to replace $d\kappa'_rd\kappa'_i$ for short notation. $L_{\kappa'_r}$ is the length for $\kappa'_r$ in the reciprocal space while $L_{\kappa'_i}$ is for $\kappa'_i$. In Eq.(\ref{delt}), $\frac{1}{L_{\kappa'_r}}\int e^{-i\kappa'_r(|\vec{R}_m-\vec{R}_0|-|\vec{R}_p-\vec{R}_0|)} d \kappa'_r$ leads to the Kronecker function $\delta_{m,p}$. In the disorder solid, there is almost zero probability for more than 1 atoms have the same distance to $\vec{R}_0$. Therefore, $\delta_{m,p}$ is a good result for the integration of $\kappa'_r$. Based on $\delta_{m,p}$, the integration of $\frac{1}{L_{\kappa'_i}}\int e^{\kappa'_i(|\vec{R}_m-\vec{R}_0|-|\vec{R}_p-\vec{R}_0|)}d\kappa'_i$ gets unit. We substitute Eq.(\ref{delt}) into Eq.(\ref{ldab1}) and we have
\begin{widetext}
\begin{align}
\label{lasteq}
\ddot{\mathcal{Q}}_{\kappa}^{\alpha}&=-\sum_l \frac{1}{\sqrt{N}}e^{-i\kappa |\vec{R}_l-\vec{R}_0|}\sum_{p,\beta}\Phi_{l,p}^{\alpha,\beta}\sum_{m}\frac{1}{L_{\kappa'_r}L_{\kappa'_i}}\int e^{-i\kappa'(|\vec{R}_m-\vec{R}_0|-|\vec{R}_p-\vec{R}_0|)}d \kappa' u_{m,\beta}\nonumber \\
&=-\sum_{\beta}\frac{1}{L_{\kappa'_r}L_{\kappa'_i}}\int\left [\sum_{l,p} e^{-i\kappa |\vec{R}_l-\vec{R}_0|}\Phi_{l,p}^{\alpha,\beta}e^{i\kappa'(|\vec{R}_p-\vec{R}_0|)}\right]\left[\frac{1}{\sqrt{N}}\sum_{m} e^{-i\kappa'(|\vec{R}_m-\vec{R}_0|)} u_{m,\beta}\right] d \kappa'\nonumber \\
&=-\sum_{\beta} \frac{1}{L_{\kappa'_r}L_{\kappa'_i}}\int \mathcal{F}_{\kappa, \kappa'}^{\alpha, \beta} \mathcal{Q}_{\kappa'}^{\beta} d \kappa'.
\end{align}
\end{widetext}
We define $\mathcal{F}_{\kappa, \kappa'}^{\alpha, \beta}$ for the first bracket and use the notation of $\mathcal{Q}$ for the second bracket on the second line of Eq.(\ref{lasteq}). Then we recover Eq.(\ref{IEequ}).\\

\begin{thebibliography}{99}
\bibitem{born}
M. Born and K. Huang, \textit{Dynamical Theory of Crystal Lattices}, Oxford University Press, Oxford, 1954.
\bibitem{ziman}
J.M. Ziman, \textit{Electrons and Phonons},  Oxford University Press, Oxford, 1960.
\bibitem{dove}
M. T. Dove, \textit{Introduction to Lattice Dynamics}, Cambridge University Press, Cambridge, 1993.
\bibitem{henry1}
Hamid Reza Seyf, Luke Yates, Thomas L. Bougher, Samuel Graham, Baratunde A. Cola, Theeradetch Detchprohm, Mi-Hee Ji, Jeomoh Kim, Russell Dupuis, Wei Lv and Asegun Henry, \textit{Rethinking phonons: The issue of disorder}, npj Comput Mater, {\bf 3}, 49 (2017).
\bibitem{henry}
Freddy DeAngelis, Murali Gopal Muraleedharan, Jaeyun Moon, Hamid Reza Seyf,
Austin J. Minnich, Alan J. H. McGaughey, and Asegun Henry, \textit{Thermal Transport in Disordered Materials}, Nanosc Microsc Thermophys. Eng., {\bf 23 (2)}, 81 (2019).
\bibitem{henry2}
W. Lv and A. Henry, \textit{Examining the validity of the phonon gas model in amorphous materials}, Sci. Rep., {\bf 6}, 37675 (2016).
\bibitem{phononfail1}
Philip B. Allen, Xiaoqun Du, Laszlo Mihaly, and Laszlo Forro, \textit{Thermal conductivity of insulating $Bi_2Sr_2YCu_2O_8$ and superconducting $Bi_2Sr_2CaCu_2O_8$: failure of the phonon-gas picture}, Physical Review B, {\bf 49}, 9073 (1994).
\bibitem{phononfail2} 
T. Sun and P. B. Allen, \textit{Lattice thermal conductivity: computations and theory of the high-temperature
breakdown of the phonon-gas model}, Physical Review B, {\bf 82}, 224305 (2010).
\bibitem{MDFrenkel}
Daan Frenkel,Berend Smit, \textit{Understanding Molecular Simulation: From Algorithms to Applications}, Academic Press,  San Diego, 2002.
\bibitem{MD1}
A. J. McGaughey and M. Kaviany, \textit{Phonon transport in molecular dynamics simulations: formulation and
thermal conductivity prediction}, Advances in Heat Transfer, {\bf 39}, 169 (2006).
\bibitem{MD2}
P. K. Schelling, S. R. Phillpot, and P. Keblinski, \textit{Comparison of atomic-level simulation methods for computing
thermal conductivity}, Physical Review B, {\bf 65}, 144306 (2002).
\bibitem{MD3}
R. J. Hardy, \textit{Energy-flux operator for a lattice}, Phys. Rev., {\bf 132}, 168(1963).
\bibitem{LD1}
P. B. Allen and J. L. Feldman, \textit{Thermal conductivity of disordered harmonic solids}, Physical Review B, {\bf 48}, 12581 (1993).
\bibitem{LD2}
Philip B. Allen, Joseph L. Feldman, Jaroslav Fabian and
Frederick Wooten, \textit{Diffusons, locons and propagons: character of atomie yibrations in amorphous Si},
Philosophical Magazine B, {\bf 79}, 1715 (1999).
\bibitem{LD3}
Joseph L. Feldman, Mark D. Kluge, Philip B.Allen and Frederick Wooten, \textit{Thermal conductivity and localization in glasses: Numerical study of a model of amorphous silicon},
Physical Review B, {\bf 48}, 12589 (1993).
\bibitem{LD4}
H. R. Seyf and A. Henry, \textit{A method for distinguishing between propagons, diffusions, and locons}, J Appl Phys, {\bf 120}, 025101 (2016).
\bibitem{moon1}
J. Moon and A. J. Minnich, \textit{Sub-amorphous thermal conductivity in amorphous heterogeneous nanocomposites}, RSC Adv, {\bf 6}, 105154 (2016).
\bibitem{moon2}
J. Moon, B. Latour, and A. J. Minnich, \textit{Propagating elastic vibrations dominate thermal conduction in
amorphous silicon}, Physical Review B, {\bf 97}, 024201 (2018).
\bibitem{lmps1}
S. Plimpton, \textit{Fast Parallel Algorithms for Short-Range Molecular Dynamics}, J Comp Phys, {\bf 117}, 1 (1995).
\bibitem{lmps2}
http://lammps.sandia.gov.
\bibitem{sw}
F. H. Stillinger and T. A. Weber, \textit{Computer simulation of local order in condensed phases of silicon}, Physical review B, {\bf 31}, 5262 (1985).
\bibitem{zal}
R. Zallen, \textit{The physics of amorphous solids}, Wiley,
New York, 1983.
\end{thebibliography}

\end{document}